\documentclass[11pt]{article}

\usepackage{tikz}
\usepackage{pgfplots}

\pgfplotstableread[row sep=\\,col sep=&]{
   interval & Pol1 & Pol2 & Pol3 \\
  n=24     & 8  & 6 & 4   \\
 n=32     & 6 & 6  & 4  \\
 n=64    & 6 & 4 & 4 \\
  n=96   & 6 & 4 & 4 \\
  n=128   & 4  & 4 & 4 \\
  }\mydata

\begin{document}

\title{CRC selection for decoding of CRC-polar concatenated codes}

\author{Tsonka Baicheva* and Peter Kazakov
\thanks{Tsonka Baicheva is with Institute of Mathematics and Informatics, Bulgarian Academy of Sciences, 8 N. Gabrovski Str., 5000 Veliko Tarnovo, Bulgaria.
        {\tt\small tsonka@math.bas.bg}}%
\thanks{Peter Kazakov is with Institute of Mathematics and Informatics, Bulgarian Academy of Sciences, 8 N. Gabrovski Str., 5000 Veliko Tarnovo, Bulgaria and Skyscanner (Bulgaria) LTD, 2 Positano Square, 1000 Sofia, Bulgaria
        {\tt\small peter.kazakov@gmail.com}}%
				}
         
\maketitle

\begin{abstract}
An efficient scheme to increase the performance of polar codes at short and moderate block lengths is a concatenation of CRC code and a polar code. In order to obtain better result of the concatenation, a CRC code with best error control performance among all CRC codes with a fixed number of check bits has to be used. In this work we investigate CRC codes of 11 to 19 parity bits and determine those of them which have maximum minimum distance at any length it can be used. For CRC codes of 24 parity bits we were not able to perform complete search and we present the best obtained results. The investigation shows that there are better CRC polynomials of degrees 11, 16 and 24 than those suggested by the 3rd Generation Partnership Project (3GPP).
\end{abstract}

Keywords: CRC codes, polar codes, decoding of CRC-polar concatenated codes.

\section{Introduction}
A class of codes called {\it optimized codes for bitwise multistage decoding} were
constructed  by  Stolte in 2002 \cite{ST}. These codes did not get attention during the years as there was no explicit proof of their capacity achieving property. In 2009 Ar{\i}kan \cite{A} introduced polar codes as a new form of forward error correction and proved that they can achieve the capacity of any binary memoryless symmetric
(BMS) channel with efficient encoding and successive cancellation (SC) decoding. More precisely, Ar{\i}kan proved that they can achieve the capacity of any discrete binary memoryless symmetric channel when the code length $n$ approaches infinity. However, when these codes are used at short and moderate block lengths their performance is unsatisfactory. Tal and Vardy \cite{TV2011,TV2012} showed that there are two main reasons for this. The first one is that polar codes are inherently weak at short to moderate block lengths. The second reason is that the performance of the SC decoder is significantly degraded compared to that of the maximum-likelihood (ML) decoder. To improve the performance of the SC decoder at short and moderate block lengths they proposed a successive cancellation list (SCL) decoding algorithm, which performs almost as good as the ML decoders when the list size $L$ is large. The algorithm considers up to $L$ decoding paths concurrently at each decoding stage. Then a single codeword is selected from the list as output. If the most likely codeword is selected, simulation results show that the resulting performance is very close to that of an ML decoder, even for moderate values of $L$. 

The performance of polar codes, however, even under the ML decoding scheme is inferior to that of low-density parity-check (LDPC) codes. To strengthen the performance of polar codes, Tal and Vardy \cite{TV2011,TV2012} suggested to use a "genie" to predict what codeword in the list was the transmitted one if it is presented in the list. To implement such a genie a concatenation scheme of cyclic redundancy check (CRC) codes and polar codes, named CRC-aided polar codes, was found to be effective. It has been shown that CRC-aided polar codes under SCL decoding are capable of achieving better performance than those of LDPC and turbo codes \cite{TV2011,TV2012, LST}.  

Thanks to their excellent error control performance and code rate flexibility,  the family of CRC-aided polar codes has been adopted as a channel code for the control channel of the 5G network of the 3rd generation partnership project \cite{PolarCRC}. However, the decoder implementation is one of the big practical problems and a low complexity realization of the SCL decoder is of great importance. Resent research results suggest reducing the decoding complexity by efficiently managing decoding paths. An adaptive SCL decoder for CRC-aided polar codes was proposed in \cite{LST}.  A unified description of successive cancellation decoding and its improved version with a list or stack is provided in \cite{NC} and the CRC-aided successive cancellation list/stack (CA-SCL/SCS) decoding schemes are proposed. Some studies have used multiple CRC codes to address different complexity issues of the decoding \cite{GLL}-\cite{KKJK}. Simulation results including application of different CRC codes for CRC-aided list decoding are obtained by Wasserman \cite{W}. A heuristic construction of polar codes for SCL decoding to optimize the frame error rate for a given list size is proposed in \cite{YPBIBX}. 

As a further improvement of the performance of the ML decoding, the CRC-aided sphere decoding (CA-SD) algorithm is proposed in \cite{PDN, YPLT}. The CA-SD can provide better performance than the CA-SCL decoding and leads to stable performance over a wide range of code rates.

In all of the above cited works authors do not explain how they choose the applied CRC codes or use simulations of particular schemes to make the choice. It is also not clear how the suggested for the enhanced Mobile Broadband (eMBB) service category in the 5-th generation wireless systems (5G) \cite{PolarCRC} CRC codes are chosen. 

In this work we briefly recall the basic facts about the error detection performance of CRC codes and apply a complete search on all CRC codes with $11,\ldots,19$ parity bits for the most often used by CRC-aided polar codes lengths $M=512, 1024, 2048, 4096$. This way we find the optimal (having maximum minimum distance) codes for each of the considered number of parity bits and lengths. For CRC codes of 24 parity bits we perform a complete search for code length $M=512$. For the lengths $M=1024, 2048, 4096, 8192$ we present the best obtained results, but do not claim that they are optimal. The results are obtained by a computer search using a binary search algorithm with different pruning techniques.  

We do not consider CRC codes of less than 11 parity bits because in \cite{Bay} all the necessary data to evaluate the error control performance of these codes has been calculated. 
 
\section{CRC codes for error detection}

We suggest Peterson and Weldon's \cite{PW} and Lin and Costelo's \cite{LC} books as a very good sources on CRC codes, their description and their properties. We will recall only some basic facts about the structure of CRC codes.
 
Cyclic redundancy-check codes (CRC codes) are shortened binary cyclic codes. They are linear codes where each codeword $c=[i, r]$ consists of $n=k+p$ binary digits and is obtained by adding in a definite way a block $$r=[r_0, r_1, \dots, r_{p-1}]$$ of $p$ parity bits to an information block of $k$ binary digits $$i=[i_0,i_1, \dots,i_{k-1}].$$ 

It is convenient for our work to consider binary information in terms of polynomials. Thus instead of the information block of $k$ binary digits, we will consider the polynomial $$i(x)=i_0+i_1x+\dots+i_{k-1}x^{k-1},$$ and instead of the block of $p$ parity bits the polynomial $$r(x)=r_0+r_1x+\dots+r_{p-1}x^{p-1}.$$ A CRC code of $p$ check bits is defined by its {\it generator polynomial} $g(x)$ of degree $p$ whose leading and zero coefficients are nonzero. The block $r(x)$ of parity bits is computed from the information bits $i(x)$ in such a way that $$r(x)\equiv (x^p.i(x))~mod~g(x).$$ Then the codeword is $$c(x)=x^pi(x)+r(x)=q(x)g(x)+r(x)+r(x)=q(x)g(x).$$ According to this coding all the information symbols are before the check symbols in the codeword, thus the encoding is systematic. All codewords of the code are divisible by $g(x)$ and all polynomials divisible by $g(x)$ are code polynomials. Therefore, to detect errors at the receiving end, we only have to check if the received word is divisible by $g(x)$, i.e. is a codeword.

We will note that $g(x)$ is a polynomial dividing $x^{n_c}+1$, where $$n_c=\min \{m\vert x^m\equiv 1~mod~g(x)\}.$$ The number $n_c$ is called {\it order of the polynomial} $g(x)$. So, we consider the binary cyclic $[n_c,n_c-p]$ code $D$ generated by the polynomial $g(x)$ of degree $p$, where $n_c=x^m+1$.

In practice, we often use some $[n,n-p]$ subcode $C$ (a CRC code) of the code $D$ obtained by shortening $D$ in $n_c-n$ positions. In general, the CRC code is not cyclic, but has at least the same error-correcting capabilities as the code from which it is derived and its encoding and decoding can be done in the same way as for the original cyclic code.

Let us denote by $d(C_{g,n})$ the minimum distance of an $[n,n-p]$ CRC code generated by the polynomial $g(x)$ and shortened in $n_c-n$ positions. Any $[n,n-p]$ CRC code of minimum distance $d(C_{g,n})$ can detect up to $d(C_{g,n})-1$ errors. That is why, codes having maximum minimum distance for a fixed length $n$ are of practical interest. 

Having this in mind, we formulate an optimization criterion for our search and determine which CRC codes are best with respect to it.  

\section{Optimization criterion}

We accept that information packages transmitted by a particular polar code may change their lengths during the communication depending on the channel conditions. For instance, if channel conditions degrade, one may reduce the number of information bits, but still use the same CRC code. The number of information bits sent via one codeword is also dependent on the characteristic of the information stream which has to be transmitted and may vary during the communication. That is why our criterion is chosen to be suitable for polar codes with information packages whose length may vary. We aim to find optimal CRC polynomials for a fixed number of parity bits $p=11,\dots, 19$ and $24$ and lengths between $p+1$ and $M$, where $M=512, 1024, 2048, 4096$ and $8192$. 

Therefore for any polynomial $g(x)$ of degree $p$, which can be a generator polynomial of a CRC code we calculate

$$S_g(M) = \sum_{n=p+1}^{M}d(C_{g, n})$$

and find the codes having maximum $S_g(M)$. 

It is proved \cite{V97} that the computation of the minimum distance of a linear code is an NP-complete problem. This means that we do not have a linear time algorithm for solving this problem. More precisely, for a fixed length $n$, we have to generate all $2^k$ codewords and to determine their weights in order to find the smallest one as the minimum distance of a binary linear code is equal to the minimum weight of its codewords. We have to repeat this procedure $M-p$ times to obtain $S_g(M)$ for a particular CRC code.

It is clear that we have to perform very heavy computations for the considered in this work CRC codes. In order to make the calculations feasible, we use some pruning techniques which are described in the next section.

\section{Method of investigation}

We investigate CRC codes generated by polynomials of degrees $p=11,\dots, 19$ and $24$ of predefined maximal length $L$. This means that the dual codes of the investigated by us codes have dimensions 11, ...,19 and 24, i.e. exactly $2^{11},\dots,2^{19}$ and $2^{24}$ codewords for any of the considered code lengths $(p+1\leq n\leq L)$. Then, instead of generating all $2^{n-p}$ codewords of the CRC code of length $n$ for $(p+1\leq n\leq L)$, we generate only $2^p$ codewords of length $n$ of its dual code for $(p+1\leq n\leq L)$. Then we compute the minimum distance of the corresponding CRC code in linear time via MacWilliams' identities \cite{MacW}.

We consider only polynomials which are not divisible by $x$ because instead of $g(x)=xg_1(x)$ we can always use $g_1(x)$ to obtain the same results. Therefore, we have to consider $2^{p-1}$ polynomials for each $p$. We additionally reduce the number of checked polynomials by omitting the reciprocal ones as they generate equivalent codes.

One important characteristic of each polynomial to be used as a generator polynomial of a CRC code is its order $n_c$ as this is the length of the cyclic code $D$. Codes of lengths $n > n_c$ are repetition codes and their minimum distance is 2 (see \cite{Bay}, section V). Therefore, for each particular polynomial, we are interested only in codes of lengths up to $n_c$ and do not consider polynomials of order $n_c$ smaller than the maximum length $L$ for which a CRC code is used.

Then we repeat the following steps:

1. Determine $d(C_{g, p+1})$, $d(C_{g, N})$ and $d(C_{g, M})$ for some arbitrary chosen length $N$ between $p+1$ and $M$ but close to $p+1$. 

2a. If $d(C_{g, N})$=$d(C_{g, p+1})$ or $d(C_{g, N})$=$d(C_{g, M})$, we do not need to calculate the values of $d(C_{g, m})$ for  $m\in [p+2, \dots, N-1]$ or $m\in [N+1,\dots, M]$ respectively, as the minimum distance is an integer number and all values of $d(C_{g, m})$ for the corresponding interval are equal to $d(C_{g, N})$.

2b. Otherwise we apply a binary split to the intervals $[p+2, \dots, N-1]$ and $[N+1,\dots, M-1]$.

3. We repeat steps 2a and 2b until all values of $d(C_{g, m})$ for $m\in [p+1, \dots, M]$ are determined.

This procedure speeds up significantly the process of determination of the minimum distance of the investigated codes because we do not have to calculate the minimum distance for all shortened lengths $p+1,\dots, M-1$. The reason is that if $d(C_{g, m})=d(C_{g, m+t})$ for any positive $t$, then we will have the same minimum distance for the whole interval between $m$ and $m+t$. 

We apply the above algorithm to determine the values of $S_g(M)$ for all CRC codes with a fixed number of parity bits $p$ ($p=11,\dots, 19$ and $24$). Then we determine the CRC codes having maximum $S_g(M)$ for the given $p$.

\section{Results}

By the suggested method of investigation, we determine values of $S_g(M)$ for all CRC codes with generator polynomials of degrees $p=16, \ldots, 19$ and $24$ and for four values of $M=512, 1024, 2048, 4096$, as these are the most often used lengths of CRC-aided polar codes with SCL decoding. For a completeness of the research, we also include results for $p=11,\dots,15$, $M=512$, $p=16, 24$ and $M=8192$.

In Tables 1-4 we show only the best according to our optimization criterion polynomials, their orders and values of $S_g(M)$. To present the information in a more compact way, we use hexadecimal notation of the polynomials, i.e. for $x^{16}+x^7+x^6+x^5+x^4+x+1$ we use $100f3$. We mark in bold the polynomials with maximum value of $S_g(M)$ for a given $p$ and these are our recommendations for best CRC polynomials. We put a dash in the column with values of $S_g(M)$ when the order of the polynomial is less than $M$.

We recall that the reciprocal polynomials generate equivalent codes, thus the reciprocal of the presented polynomials can also be used.

For comparison, we include in Tables 1, 2 and 4 the CRC polynomials suggested in \cite{PolarCRC} and \cite{LDW} and the two standardized (IBM and CCITT) CRC codes of 16 bit redundancy. The authors of \cite{LDW} used a slightly different notation for the generator polynomials. They present them without the least significant bit (which is always 1), so we have converted their notation to our.  

\subsection{Results for $p=11,\dots, 15$}

Only CRC codes of $M=512$ are investigated and the best polynomials are given in Table 1. For $p=12$ and $p=15$ we suggest two polynomials as the values of $S_g(M)$ are very close.
  
\begin{table}
\caption{Best CRC codes with up to 15 check bits}
\label{table_example1}
\begin{center}
\begin{tabular}{|c|c|c|c|}
\hline
Polynomial $g$&Degree&Order&$S_g(512)$\\
\hline
\textbf{93f}& 11 & 762 &	2044		\\
a0f \cite{LDW} & 11 & 146 & - \\
e21 \cite{PolarCRC} & 11 & 2047 & 1565 \\
\textbf{1957}& 12 & 1778 & 2056  \\
\textbf{1637}& 12 & 1905 & 2054  \\
1421 \cite{LDW} & 12 & 2047 & 2000 \\
\textbf{299d} & 13 &	3556 &	2078 \\
3c1f \cite{LDW} & 13 & 8191 & 1612 \\
\textbf{6e57} & 14 &	8190 & 2116 \\
629d \cite{LDW} & 14 & 1016 & 2034 \\
\textbf{86ef} &	15 & 15748	& 2138	\\
\textbf{9be5}& 15	& 16383	& 2134	\\
c099 \cite{LDW} & 15 & 5355 & 2044 \\
\hline
\end{tabular}
\end{center}
\end{table}

\subsection{Results for p=16}

As can be seen from Table 2 two polynomials have maximum $S_g(M)$ for all four values of $M=512, 1024, 2048, 4096$.

\begin{table}
\caption{Best CRC codes with 16 check bits}
\label{table_example2}
\begin{center}
\begin{tabular}{|c|c|c|c|c|c|c|}
\hline
Polynomial&Order&512&1024&2048&4096&8192\\
\hline
158ff& 7161 & 2196 & 4244	& 8340	& 16532 & -	\\
\textbf{1a2eb}&32767 & 2196 & 4244  & 8340  & 16532 & 32916 \\
11cc3 \cite{LDW} & 1905 & 2080 & 4128  & -  & - & - \\
18005 (IBM) & 32767 & 1984 & 4032  & 8128  & 16320 & 32704 \\
11021 (CCITT) \cite{PolarCRC} & 32767 & 1984 & 4032  & 8128  & 16320 & 32704 \\
\hline
\end{tabular}
\end{center}
\end{table}

The polynomial $158ff$ (and its reciprocal $1fe35$) has $d=4$ for $n=112 \ldots  7161$, $d=6$ for $n=26 \ldots  111$, $d=8$ for $n=18 \ldots 25$, $d=12$ for $n=17$. The polynomial $1a2eb$ (and its reciprocal  $1ae8b$) has $d=4$ for $n=110 \ldots 32767$, $d=6$ for $n=28 \ldots 109$, $d=8$ for $n=19 \ldots  27$, $d=10$ for $n=17,18$. 

We also include in Table \ref{table_example2} the standardized IBM polynomial which has the same order as $1a2eb$.  

We have additionally checked that the polynomial $1a2eb$ has maximum $S_g(M)=32916$ for $M=8192$. 

Fig. 1 illustrates that the two standardized codes 18005 and 11021 have minimum distance 4 for the whole interval of code lengths between 24 and 128.
The CRC code generated by the $1a2eb$ polynomial keeps its high minimum distance for a long range of short code lengths which determines The two standardized codes 18005 and 11021 have minimum distance 4 for the whole interval of code lengths between 24 and 128.

\begin {figure}
\begin {center}
   \begin{tikzpicture}
    \begin{axis}[
            ybar,
            bar width=.2cm,
            width=0.5*\textwidth,
            height=0.5*\textwidth,
            legend style={at={(0.5,1)},
                anchor=north,legend columns=-1},
            symbolic x coords={n=24,n=32,n=64,n=96,n=128},
            xtick=data,
            nodes near coords,
            nodes near coords align={vertical},
            ymin=0,ymax=10
        ]
        \addplot table[x=interval,y=Pol1]{\mydata};
        \addplot table[x=interval,y=Pol2]{\mydata};
        \addplot table[x=interval,y=Pol3]{\mydata};
        \legend{1a2eb, 11cc3, (18005,11021)}
    \end{axis}
\end{tikzpicture}
\caption{Comparison between the best CRC codes presented in Table 2.}
\end {center}
\end {figure}
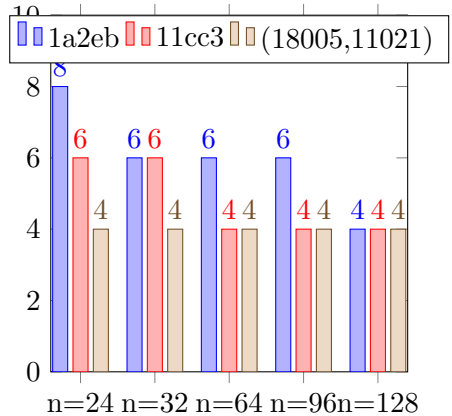

\subsection{Results for p=17, 18, 19}

The results from Table 3 show that some of the codes are optimal up to a given length, but cannot be used for greater lengths, because the order of their generator polynomials is smaller. Thus, if we need codes for longer lengths, it is better to chose among the included in the table polynomials of greater order. 

\begin{table}
\caption{Best CRC codes with 17,18,19 check bits}
\label{table_example3}
\begin{center}
\begin{tabular}{|c|c|c|c|c|c|c|}
\hline
Polynomial&Dgr&Order&512&1024&2048&4096\\
\hline
\textbf{2ca6d}& 17 & 57316 & \textbf{2264} &	\textbf{4312}&	\textbf{8408}&	\textbf{16600}\\
658d3& 18 &	514	& \textbf{2502} & - & - & - \\
\textbf{446b7} & 18 & 42966 & 2358 &	\textbf{4406}&	\textbf{8502}& \textbf{16694} \\
ad0b5 & 19 & 513 & \textbf{2994} & - & - & - \\
ae975 &	19 & 1028 &	2514 & \textbf{4562} & - & - \\
\textbf{c492f} & 19 & 81915 & 2472 & 4520 & \textbf{8616} & 	\textbf{16808} \\

\hline

\end{tabular}
\end{center}
\end{table}

\subsection{Results for p=24}

For this number of parity bits we were able to obtain optimal results by the suggested method only for $M=512$. For greater $M$, we do not perform complete search and the best obtained results are given in Table 4. Thus, results from this section are not claimed to be optimal except for $M=512$, but they are good enough to be used in practical applications.

It can be seen that there is no polynomial which is best for all $M$. For different $M$ different polynomials have to be used. Our calculations confirm that the polynomial chosen for the FlexRay standard \cite{Flex} is a good choice for code lengths $n \leq 4094$. 

\begin{table}
\caption{Suggested CRC codes with 24 parity bits}
\label{table_example4}
\begin{center}
\begin{tabular}{|c|c|c|c|c|c|c|}
\hline
Polynomial&Order&512&1024&2048&4096&8192\\
\hline
\textbf{11175b7}	& 1195740&	\textbf{3134}&	5330&	9426&	17618&	34002 \\
15d6dcb \cite{Flex} & 4094	& 3116 & \textbf{6188} & \textbf{12332}	& -	& - \\
1eb83af	& 4098	& 3014	& 6086	& 12230	& \textbf{20426} & - \\

\textbf{1ce467f} &	8355585	& 3042 &	5708 & 9804	& 17996	& \textbf{34380} \\

1864cfb \cite{PolarCRC} & 8388607 & 3014 & 5120 & 9216	& 17408	&  33792 \\
1800063 \cite {PolarCRC} & 8388607 & 1960 & 4008 & 8104	& 16296	&  32680 \\
1b2b017 \cite {PolarCRC} & 1168146 & 2366 & 4414 & 8510	& 16702	&  33086 \\
\hline

\end{tabular}
\end{center}
\end{table}

\section{Conclusion}
Our investigations concern CRC codes which can be used in CRC-aided polar codes with SCL or SD decoding. We propose an optimization criterion and find optimal according to this criterion CRC codes of $11,\ldots, 19$ parity bits and different lengths. For CRC codes of $24$ parity bits and lengths greater than 512 the search space is enormous and we were not able to perform a complete search. In that case we present the best obtained results. We show that the FlexRay standard uses the best know CRC code with 24 parity bits for lengths 1024 and 2048. We also show that there are better CRC polynomials of degrees 11, 16 and 24 than those suggested in \cite{PolarCRC}.

CRC codes are the second most widely used codes for error detection after the simple parity check. They are applied in a big variety of coding schemes for data transmission and storage. The results obtained in this work can be used in any code scheme where the length of the information block may vary up to $M-p$.

\section{Acknowledgments}
This work was partially supported by the Bulgarian National Science Fund under Grant No. 12/8, 15.12.2017.

 \bibliographystyle{ACM-Reference-Format}

\end{document}